\documentclass{article}
\usepackage{spconf,amsmath,amssymb,graphicx,hyperref}
\usepackage{booktabs}
\usepackage{multirow}
\usepackage{xcolor}

\newcommand{\myparagraph}[1]{\noindent\textbf{#1}}
\title{BaldWhisper: Faster Whisper with Head Shearing and Layer Merging}
\name{Yaya Sy, Christophe Cerisara, Irina Illina\thanks{This project was provided with computing HPC and storage resources by GENCI at IDRIS thanks to the grant 2025-AD011011668R5 and AD011014953 on the supercomputer Jean Zay’s A100 partition}}
\address{LORIA, CNRS, Nancy, France}
\begin{document}
%
\maketitle
\begin{abstract}
Pruning large pre-trained transformers in a data-scarce scenario is challenging, as it often requires massive retraining data to recover performance. For instance, Distill-Whisper prunes Whisper by 40\% and retrains on 21,000 hours of speech, far beyond what is available for most languages. Can Whisper be made lighter and faster for edge devices in data-scarce settings? Focusing on Bambara with only 32h of speech-to-text data, we propose a new pruning recipe. Instead of vocabulary pruning, which is unsuitable due to frequent code-switching by Bambara speakers, we compress the embeddings with low-rank decomposition and feature distillation. Rather than removing layers, we merge them to limit performance loss. The final model preserves 90\% of the original performance while being 48\% smaller and 2.15x faster on a MacBook Air M1.
\end{abstract}
\section{Introduction}
One approach for local edge-device inference for speech recognition is to prune a large, pre-trained, high-performing model into a small model. However, traditional pruning methods often require substantial amounts of retraining data. For example, Distill-Whisper \cite{distill-whisper} prunes the English-only version of Whisper \cite{whisper} and further retrained on 21k hours of ASR data, a quantity not available for many languages. This leads us to our central research question: \emph{How to compress a pre-trained ASR model (Whisper) when only a small quantity of supervised data is available?}
\begin{figure}[ht]
    \centering
    \includegraphics[width=0.5\textwidth]{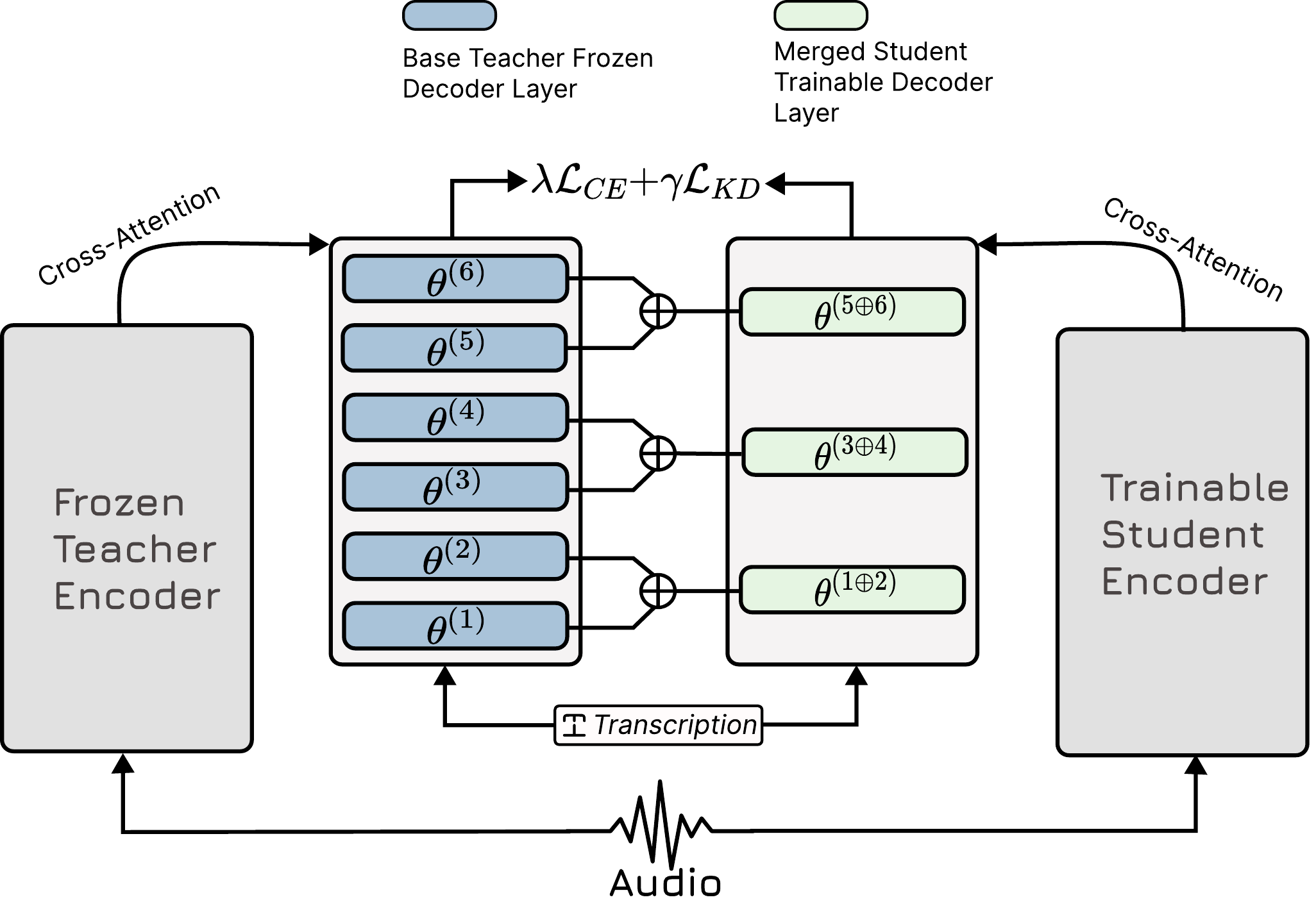}
    \caption{Layer merging in the Whisper decoder: each student layer is a weighted average of two consecutive teacher layers, then trained with Cross-Entropy + Knowledge Distillation.}
    \label{fig:layermerging}
\end{figure}

We are interested in edge-device inference of Whisper, a multilingual encoder-decoder speech-text model, for Bambara, a low-resource language spoken mainly in Mali. For pruning approaches to be effective in a low-resource scenario, they must preserve the performance of the base model as much as possible. To address this, we propose a two-stage pruning approach. First, we used layer merging as an alternative to simple pruning, aiming to preserve model performance. Second, because the model is intended for a single language, many features in the Whisper multilingual vocabulary matrix are not relevant. This matrix represents over 50\% of the decoder parameters in small Whisper models, so we shrink it using activation-aware low-rank decomposition. The resulting compressed model is 48\% smaller and 2.14x faster than the original Whisper when deployed offline on a single M1 chip, while still maintaining over 90\% of the base performance.
\section{Related Works}

\myparagraph{Deep Encoder, Shallow Decoder} We take inspiration from \cite{kasai2021deepencodershallowdecoder} who show that the decoder part of encoder-decoder architectures can be very short while still achieving good accuracy and being faster. To make the model lighter and faster, we mainly compress the decoder part of Whisper. However, contrary to \cite{kasai2021deepencodershallowdecoder}, which trains encoder-decoder models from scratch, we use a pre-trained Whisper and compress the decoder.\\
\\
\myparagraph{Distill-Whisper} \cite{distill-whisper} compress Whisper-large by removing 30 layers (out of 32) of the decoder. The pruned model further undergoes a retraining on 21k hours with sequence-level knowledge distillation \cite{kim2016sequencelevelknowledgedistillation}. While the resulting model performs as well as the base non-pruned model, the training data is large (\~21k hours of speech) and unavailable for many languages. Instead of removing layers, we merge them to limit the performance drop.
\\
\\
\myparagraph{Model Merging} We merge the decoder layers to make the model lighter and faster, which is also related to model merging approaches, used for combining models of different tasks \cite{wortsman2022modelsoupsaveragingweights, yadav2023tiesmergingresolvinginterferencemerging} or reducing the size of the model \cite{liu2024pruningmergingcompressingllms}. However, the approach \cite{liu2024pruningmergingcompressingllms} uses manifold learning to align activations before layer merging. Our approach is simpler as, motivated by the observation that adjacent layers produce similar activations, we simply compress the decoder by applying a weighted average of adjacent layers. We also study the effect of different weighting values and show that a high weight for the top layers produces better results.
\\
\\
\myparagraph{Vocabulary pruning.} In small multilingual models, the embedding matrix accounts for a large share of the parameters due to the large multilingual vocabulary. For instance, in Whisper-tiny, the embedding matrix represents 51\% of the total parameters. A common approach to address this is vocabulary pruning \cite{abdaoui-etal-2020-load, vocabprune}, which removes all tokens not used by the target language. However, this method is risky in the presence of out-of-vocabulary words and in code-switching scenarios, where foreign-language tokens must be generated. For example, Bambara speakers frequently mix in French in francophone regions such as Mali, or English in anglophone regions such as Gambia. To address this, we replace vocabulary pruning with a safer alternative: low-rank decomposition combined with feature distillation.
\\
\\
\myparagraph{Low-Rank Decomposition} is a common technique for dimensionality reduction. Recent work has shown that the activations of pre-trained transformers are low-rank, motivating activation-aware low-rank decomposition \cite{yu_compressing_2023, chen_drone}. In Whisper, the embedding matrix is particularly large due to its multilingual vocabulary, and we compress it using activation information rather than applying a standard SVD. However, unlike prior approaches, we train the low-rank embeddings using gradient-based feature distillation.

\section{BaldWhisper Approach}
\label{sec:method}
\begin{figure}[ht]
    \centering
    \includegraphics[width=0.5\textwidth]{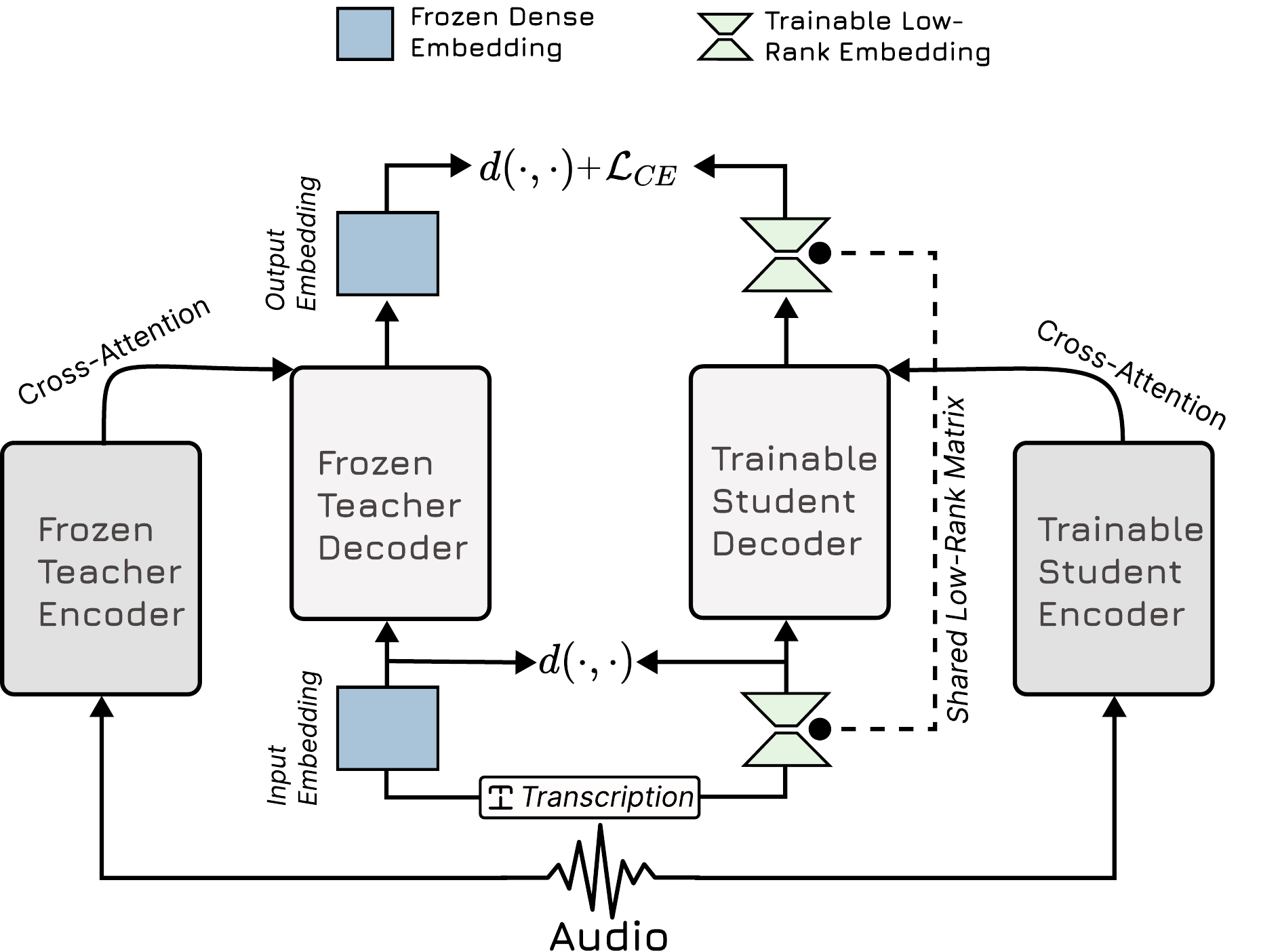}
    \caption{Head shearing of Whisper embedding via low-rank decomposition. The decomposition is activation-aware because the low-rank weights are then feature-distilled.}
    \label{fig:headshearing}
\end{figure}%
\myparagraph{Motivation.} We target offline on-device decoding. Since the encoder runs once per utterance while the decoder runs per generated token, we compress \textbf{only the decoder}. We first reduce decoder depth via layer merging, and then, using low-rank decomposition and feature distillation, we compress the shared input/output embeddings, which occopies a large fraction of decoder parameters due to the multilingual vocabulary.

\subsection{Stage 1: Layer Merging}
To accelerate inference with minimal accuracy drop, we reduce the decoder size by merging layers rather than removing them, thereby mitigating performance degradation. Let $\theta^{(l)}$ denote the parameters of the $l^{th}$ layer. Figure \ref{fig:layermerging} illustrates this stage. The decoder of Whisper-base is composed of 6 layers, then the pairs of consecutive layers are: $\{(\theta^{(1)}, \theta^{(2)}), (\theta^{(3)}, \theta^{(4)}), (\theta^{(5)}, \theta^{(6)})\}$; our approach merges each pair of layers through a weighted average:
\begin{equation}
\label{eq:merge}
\theta^{(i \oplus j)} = \frac{\alpha \cdot \theta^{(i)} + \beta \cdot \theta^{(j)}}{\alpha + \beta}
\end{equation}
{where $j$ is the $(i+1)^\text{th}$ layer, and $\alpha$ and $\beta$ are hyperparameters that control the contribution of each layer. Compressing as such Whisper-base results in a decoder with only 3 layers.
After merging, the studen model, which is composed of 6 non-compressed encoder layer and 3 merged layers, undergoes a retraining on an ASR task with a Cross Entropy loss $\mathcal{L}_{CE}$. We also distill the compressed model (student) using the non-compressed model as a teacher, since distillation has proven to be effective for speech recognition \cite{distillation, distill-whisper}. Note that only the student is trained (both the encoder and decoder) not the teacher, which is frozen.

\begin{align}
    \label{eq:hs}
    \mathcal{L}^{(1)} &= \lambda \mathcal{L}_{CE} + \gamma \mathcal{L}_{KD}
\end{align}
where $\mathcal{L}_{CE}$ is the student Cross Entropy loss for the ASR task. $\mathcal{L}_{KD}$ is the Knowledge Distillation loss computed as a standard Kullback-Leibler Divergence between the student (compressed model) and the teacher (non-compressed model).
Both losses are weighted, and $\lambda$ and $\gamma$ are tuned with a hyperparameter search.

\subsection{Stage 2: Activation-Aware Embedding Decomposition}
For small Whisper models, the input/output embedding matrix $E \in \mathbb{R}^{V \times h}$, where $V$ is the vocabulary size and $h$ the hidden size, can account for a large amount of the total parameters. Whisper, being a multilingual model, this matrix is large due to the multilingual vocabulary. However, when specializing the model to a single language, as in our case, many features of $E$ become redundant, making the embeddings compressible. This means that it is possible to further speed up the decoding by compressing the embeddings. We propose reducing the parameters in this matrix through low-rank decomposition followed by feature distillation. Figure \ref{fig:headshearing} illustrates this compression stage. We compress the shared input/output embeddings $E$ using singular value decomposition (SVD) to factorize $E \approx E_{1}E_{2}$, where $E_{1}\in\mathbb{R}^{V \times r}$ and $E_{2}\in\mathbb{R}^{r \times h}$ with $r \ll \min(V,h)$. To ensure that the low-rank embedding maximally preserves the features of the base embedding, we combine Cross Entropy loss with a Feature Distillation objective:
\begin{align}
    \label{eq:distill}
    \mathcal{L}^{(2)} &= \mathcal{L}_{CE} \nonumber \\
    &\quad + d(E[x], E_{1}[x] \cdot E_{2}) \nonumber \\
    &\quad + d(o\cdot E^{T}, (o \cdot E_{2}^{T}) \cdot E_{1}^{T})
\end{align}
where $x$ is the input sequence, $o$ the output from the decoder layer, and $d(\cdot,\cdot)$ is the Feature Distillation loss, which is defined as:

\begin{align*}
    \label{eq:distance}
    d(y, \hat{y}) &= \mathcal{L}_{\ell_{1}}(y, \hat{y}) + \mathcal{L}_{\cos}(y, \hat{y}) \\
    & = \frac{1}{h} \sum_{i=1}^h |y_i - \hat{y}_i| - \log \sigma \left( \cos \left( y, \hat{y} \right) \right) \notag
\end{align*}
where $h$ is the hidden vector dimension, $\sigma$ is the
sigmoid activation and $\cos(\cdot, \cdot)$ is the cosine similarity.
This stage further reduces the embedding parameters by $4\times$ when the rank is set from 384 to 96, with 384 being the original hidden dimension of whisper-base.

\section{Experiments}
\subsection{Method}
\myparagraph{Data.} We use 32 hours of Bambara speech data from an openly available resource\footnote{\url{https://huggingface.co/datasets/RobotsMali/jeli-asr}}. We use 50mn for development, 1h20 for testing and the remaining for training.
\\
\\
\myparagraph{Implementation Details.} We first fine-tuned the Whisper-73M parameters on the Bambara dataset for 20 epochs, on a single A100-80GB GPU, with a learning rate of $5e{-5}$. After fine-tuning, we compressed the model using the proposed method described in Section \ref{sec:method}. For the first stage, we used Bayesian optimization to find the merging hyperparameters $\alpha$ and $\beta$ that minimize the WER on the dev set. We used the implementation provided by \emph{Ax}\footnote{\url{https://ax.dev/}} for hyperparameter search and optimized for 30 \textit{iterations}. Then, the merged model undergoes another fine-tuning but using the objective in Eq~\ref{eq:hs}. We also searched, independently of the merging, for the best hyperparameters $\lambda, \gamma$ and used a learnable temperature for the Kullback-Leibler Divergence loss. For the second stage, we compressed the input/output embeddings with a rank of $r=96$, to match the size of Whisper-tiny, which represents 4x rank reduction. The model with its low-rank input/output embeddings undergoes again another fine-tuning but with the loss in Eq~\ref{eq:distill}.

\subsection{Results}

\begin{table}[ht]
    \centering
    \begin{tabular}{lccccc}
    \toprule
    \textbf{Model} & \textbf{Params (M)} & \textbf{Speed (t/s)} & \textbf{WER ($\downarrow$)} \\
    \midrule
    Whisper-base  & 73 & 66.57 & 33.11 \\
    $+$ Layer Merging & 60 & 102.24 & 34.77 \\
    $+$ Head Shearing & 38 & 142.82 & 36.49 \\
    \bottomrule
\end{tabular}
\caption{\textbf{Main results.} The effect of the Layer Merging and Head Shearing (Embedding Decomposition) on the accuracy and efficiency. The Speed was measured on an M1 MacBook Air, as the time to decode 256 tokens for each utterance in a batch of 10. The high WER may be explained by the fact that Bambara orthography is not as standardized as in English.}
\label{tab:results}
\end{table}

\myparagraph{Performance.} The main results for both compression stages are presented in Table~\ref{tab:results}. Stage~1 (Layer Merging) only represents 18\% parameter reduction compared to Whisper-base. However, after compressing the embedding, the model is \textbf{48\% smaller and retains up to 90\% of the base performance} while being compressed using only 32h of ASR training data.
\\
\\
\myparagraph{Speedup.} We benchmarked the inference speed (tokens/s) of the models on a single Macbook Air M1 by recording the wall-clock time of decoding 256 tokens for each utterance in a batch of 10. Table~\ref{tab:results} shows that the compressed model with layer merging is already 1.54x faster compared to the base model, while being only 18\% smaller. This is because the decoder of the compressed model has fewer decoder layers, which overall accelerates inference. When additionally compressing the embedding, the inference is \textbf{2.15x faster.}
\\
\\
\myparagraph{Comparison to Whisper-tiny.} For the same number of parameters, Whisper-tiny has 4 decoder layers and a huge embedding matrix, while our compressed Whisper-base has 3 decoder layers and a low-rank embedding matrix. The compressed model achieves \textbf{142.82} tokens per second while Whisper-tiny generates at a speed of \textbf{116.24} tokens per second.

\section{Analysis}
\myparagraph{How to choose the merging parameter $\alpha$ \& $\beta$?} To choose the best merging weights $\alpha$ (importance of layer 1) and $\beta$ (importance of layer 2), we performed a hyperparameter search using Bayesian optimization as implemented in \textit{Ax} library. We searched for 30 iterations by training at each time on 30\% of the training set and testing on 60\% of the development set. Figure~\ref{fig:convergence} provides the empirical motivation for our settings, demonstrating that the bottom layers ($\alpha$) should consistently contribute less to the weighted average.

\begin{figure}[ht]
    \centering
    \includegraphics[width=0.9\linewidth]{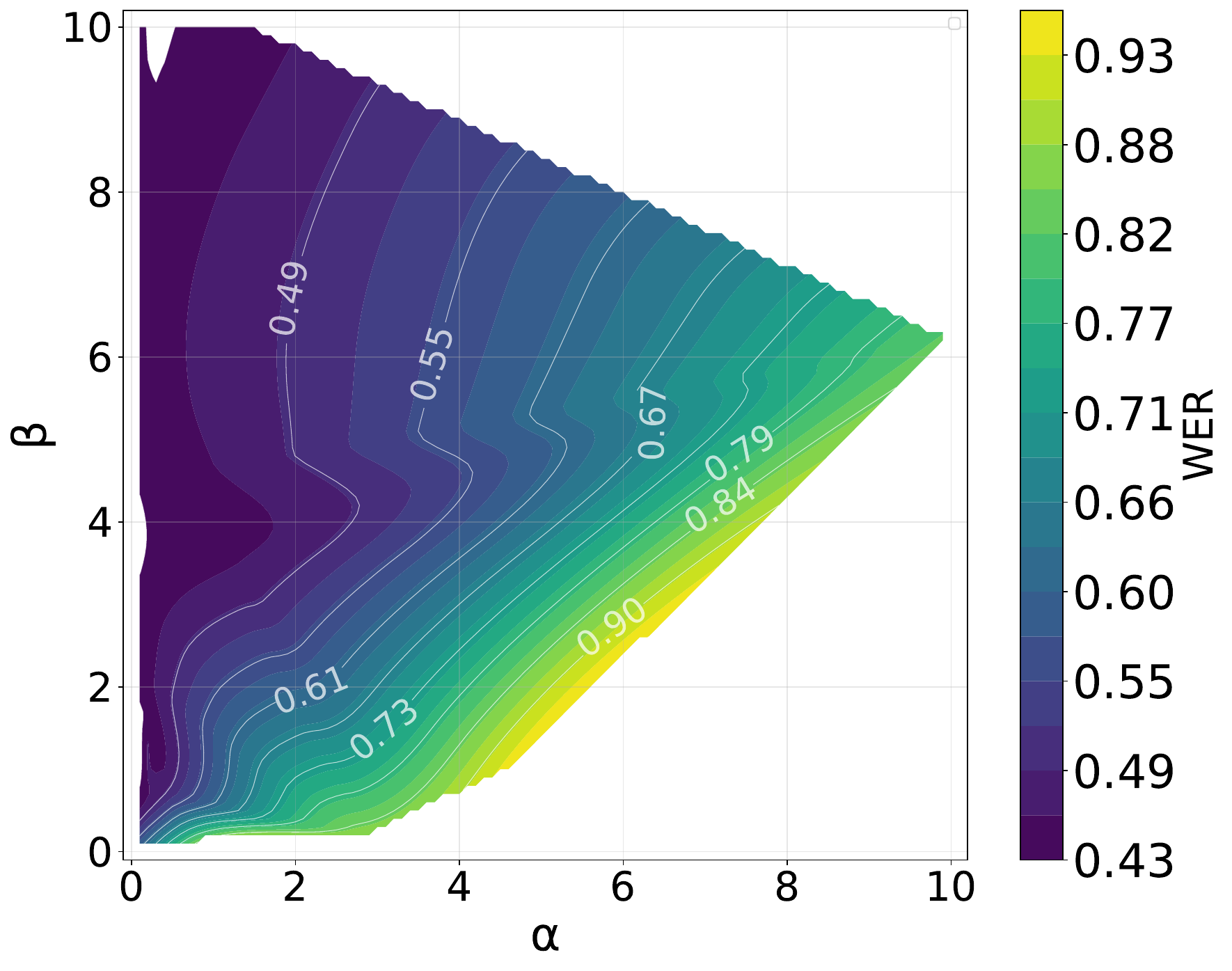}
    \caption{Choice of values $\alpha$ and $\beta$. The results of Bayesian hyperparameter search suggest that the only constraint is that $\alpha$ should be small.}
    \label{fig:convergence}
\end{figure}

\myparagraph{Why merging adjacent layers?} In the proposed approach, we merge pairs of adjacent layers, but any subset of layers may technically be merged. Exploring the optimal merging strategy certainly deserves its own set of experiments, however, because of limited computing resources and time, we rather relied on the following assumption: merging layers that share similar activations shall limit the impact on performance.
Therefore, we compare the activation similarities between all possible pair of layers of the decoder. The activations of the decoder are computed by forwarding the supervised validation corpus into Whisper-base. Figure~\ref{fig:similarities} shows the results, and we can see that adjacent layers tend to have similar activations, motivating merging pairs of adjacent layers together.

\begin{figure}[!h]
    \centering
    \includegraphics[width=0.9\linewidth]{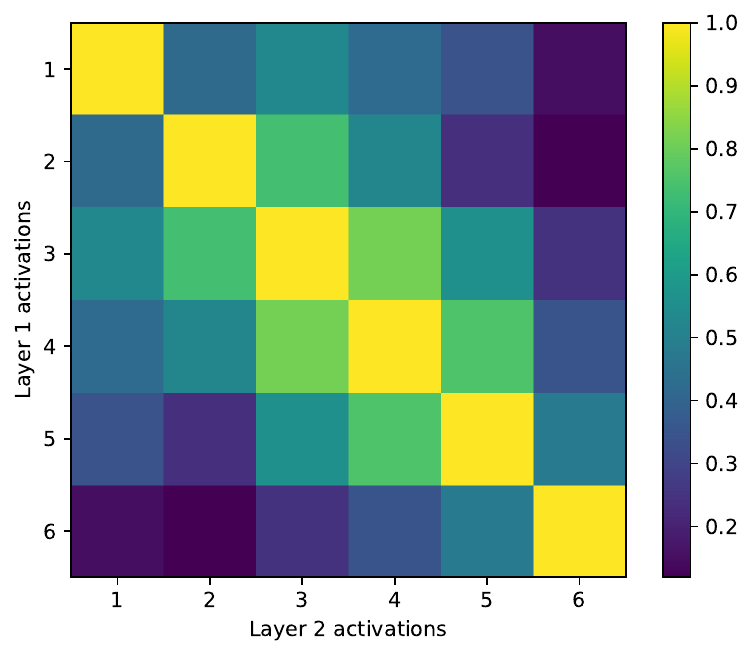}
    \caption{Visualization of the activation similarities between all possible pairs of layers of the decoder. Consecutive pairs of layers tend to be more similar, suggesting that they can be merged.}
    \label{fig:similarities}
\end{figure}
\myparagraph{Comparison to random pruning and brute force search.}
We compared layer merging to random pruning by deleting 3 layers at random and training/evaluating each pruned model on the same amount of data. On average, layer merging converges faster in WER and is more reliable, whereas random pruning shows large variance and strongly depends on which layers happen to be removed. A stronger pruning baseline would be to try all sets of $k$ layers to remove and keep the best, but this is quickly prohibitive: it requires $\binom{L}{k}$ full trainings. For example, Whisper-large has $L{=}32$ and $k{=}3$, so $\binom{32}{3}{=}4960$ runs), which is unrealistic in practice.

\myparagraph{Confidence Intervals}
At 1.54x speedup, the WER for Layer Merging (34.77\% ± 2.44\%) maintains performance statistically indistinguishable from baseline (33.11\% ± 2.41\%). Even at 2.15x speedup (36.49\% ± 2.47\%), the degradation remains modest and predictable, demonstrating robust performance across the compression range.

\section{Conclusion}
We design a new pruning approach to work in low-resource scenario and applied it to Whisper for Bambara. We show that merging adjacent layers instead of pruning limits the performance drop. The compressed model, with half of the layers and low-rank embedding, is 2.15x faster and 48\% smaller while maintaining over 90\% of the performance of the base non-pruned model. Several levers remain to further improve speed and performance, such as specializing $\alpha, \beta$ per layer and searching for their optimal values. Also, the merging function we used is a simple weighted average, but many new merging methods \cite{yadav2023tiesmergingresolvinginterferencemerging, yu2024languagemodelssupermario} may be beneficial. Since the proposed approach operates on standard Whisper Transformer blocks, we expect it to be transferable across languages and plan to investigate this in future work.

\bibliographystyle{IEEEbib}
\bibliography{strings,refs}

\end{document}